\documentstyle[11pt,moriond,epsfig]{article}

\input psfig
\begin{document}

\title{COSMIC MICROWAVE BACKGROUND OBSERVATIONS AS A WAY TO DISCRIMINATE AMONG INFLATION MODELS}

\author{ WILLIAM H. KINNEY$^{\rm a}$, SCOTT DODELSON$^{\rm a}$, EDWARD W. KOLB$^{\rm a,b}$ }

\address{$^{a}$Fermi National Accelerator Laboratory\\Batavia, IL 60510-0500\\$^{b}$ Department of Astronomy and Astrophysics, Enrico Fermi Institute, The University of Chicago, Chicago, IL 60637}

\maketitle\abstracts{
The upcoming satellite missions MAP and Planck will measure the spectrum of fluctuations in the Cosmic Microwave Background with unprecedented accuracy. We discuss the prospect of using these observations to distinguish among proposed models of inflationary cosmology. (Fermilab CONF-98/102-A)
}

\section{Introduction}

There has been great interest recently in the topic of parameter reconstruction from the Cosmic Microwave Background (CMB). Most of the emphasis in the literature is on the question of determining cosmological parameters such as the density $\Omega_0$, the Hubble constant $H_0$, or the value of the cosmological constant $\Lambda$. The purpose of this talk is to take a different approach and apply the machinery of CMB parameter estimation to the subject of inflationary model building.\cite{dodelson97} We assume generic features of the universe consistent with inflation (a flat universe, vanishing $\Lambda$), but allow parameters sensitive to inflation to vary. The goal is to determine how well we will be able to distinguish between popular inflation models using upcoming CMB experiments, in particular the all-sky satellite missions MAP and Planck.

The parameters that are of interest for inflation are the ratio of tensor to scalar fluctuation amplitudes measured at the quadrupole $r \equiv C^{\rm tensor}_2 / C^{\rm scalar}_2$, and the spectral index of scalar fluctuations $n$. Fixed parameters are the density of the universe $\Omega_0 = 1$, and the present vacuum energy $\Lambda = 0$. Other parameters are allowed to vary, and we plot the expected sensitivity of NASA's MAP satellite\cite{MAP} and the ESA's Planck Surveyor\cite{planck} as ellipses projected onto the $r\,-\,n$ plane.

\section{Parameterizing Inflation}

Inflation in its most general sense can be defined to be a period of accelerated expansion of the universe, during which the universe evolves toward homogeneity and flatness. This acceleration is typically due to the energy density of the universe being dominated by vacuum energy, with equation of state $p \simeq -\rho$. Within this broad framework, many specific models for inflation have been proposed. We limit ourselves here to models with ``normal'' gravity (i.e., general relativity) and a single order parameter for the vacuum, described by a slowly rolling scalar field $\phi$ (the {\it inflaton}). These assumptions are not overly restrictive -- most widely studied inflation models fall within this category, including Linde's ``chaotic'' inflation scenario, inflation from pseudo Nambu-Goldstone bosons (``natural'' inflation), dilaton-like models involving exponential potentials (power-law inflation), hybrid inflation, and so forth. Other models, such as Starobinsky's $R^2$ model and verions of extended inflation, can, through a suitable transformation, be viewed in terms of equivalent single-field models. 

The slow-roll approximation, in which the field evolution is dominated by drag from the cosmological expansion, is consistent if both the slope and curvature of the potential are small, $V',\ V'' \ll V$. This condition is conventionally expressed in terms of the ``slow-roll parameters'' $\epsilon$ and $\eta$
\begin{equation}
\epsilon \equiv {M_{Pl}^2 \over 4 \pi} \left({H'\left(\phi\right) \over
H\left(\phi\right)}\right) \simeq {M_{Pl}^2 \over 16 \pi}
\left({V'\left(\phi\right) \over V\left(\phi\right)}\right)^2,
\end{equation}
and
\begin{equation}
\eta\left(\phi\right) \equiv {M_{Pl}^2 \over 4 \pi} \left({H''\left(\phi\right)
\over H\left(\phi\right)}\right) \simeq {M_{Pl}^2 \over 8 \pi}
\left[{V''\left(\phi\right) \over V\left(\phi\right)} - {1 \over 2}
\left({V'\left(\phi\right) \over V\left(\phi\right)}\right)^2\right].
\end{equation}
Slow-roll is then a consistent approximation for $\epsilon,\ \eta \ll 1$. The
parameter $\epsilon$ can in fact be shown to directly parameterize the equation
of state of the scalar field, $p = -\rho \left(1 - 2/3 \epsilon\right)$, so
that the condition for inflation of accelerating expansion is exactly equivalent to $\epsilon < 1$. To match the observed degree of flatness and homogeneity is the universe, we require many e-folds of inflation, typically $N \simeq 50$. 

Inflation not only explains the high degree of large-scale homogeneity in
the universe, but also provides a mechanism for explaining the observed {\em
inhomogeneity} as well. The metric perturbations created during inflation are of two types:
scalar, or {\it curvature} perturbations, which couple to the stress-energy of
matter in the universe and form the ``seeds'' for structure formation, and
tensor, or gravitational wave perturbations, which do not couple to matter.
Both scalar and tensor perturbations contribute to CMB anisotropy. Scalar fluctuations can be quantitatively described by perturbations $P_{\cal R}$ in the intrinsic
curvature scalar
\begin{equation}
P_{\cal R}^{1/2}\left(k\right) = {1 \over \sqrt{\pi}} {H \over M_{Pl}
\sqrt{\epsilon}}\Biggr|_{k^{-1} = d_H}.
\end{equation}
The fluctuation power is in general a function of wavenumber $k$, and is
evaluated when a given mode crosses outside the horizon during inflation
$k^{-1} = d_H$. Outside the horizon, modes do not evolve, so the amplitude of
the mode when it crosses back {\em inside} the horizon during a later radiation
or matter dominated epoch is just its value when it left the horizon during
inflation. The {\em spectral index} $n$ is defined by assuming an
approximately power-law form for $P_{\cal R}$ with
\begin{equation}
n - 1 \equiv {d\ln\left(P_{\cal R}\right) \over d\ln\left(k\right)} \simeq 1 - 4 \epsilon + 2 \eta,
\end{equation}
so that a scale-invariant spectrum, in which modes have constant amplitude at
horizon crossing, is characterized by $n = 1$. Instead of specifying
the fluctuation amplitude directly as a function of $k$, it is often convenient
to specify it as a function of the number of e-folds $N$ before the end of
inflation at which a mode crossed outside the horizon. Scales of interest for
current measurements of CMB anisotropy crossed outside the horizon at $N \simeq
50$, so that $P_{\cal R}$ is conventionally evaluated at $P_{\cal R}\left({N =
50}\right)$. Similarly, the power spectrum of tensor fluctuation modes is given
by
\begin{equation}
P_{T}^{1/2}\left(k_N\right) = {1 \over \sqrt{\pi}} {H \over M_{Pl}}\Biggr|_{N =
50}.
\end{equation}
The tensor spectral index is
\begin{equation}
n_{T} = - 2 \epsilon = -2 {P_{\cal R} \over P_{T}}.\label{consistency}
\end{equation}
Note that $n_{T}$ is {\it not} independent of the other parameters $P_{\cal R}$, $P_{T}$, and $n$. Equation (\ref{consistency}) is known as the {\it consistency condition}, and is true only for single-field models. (In models with multiple degrees of freedom, Eq. (\ref{consistency}) weakens to an inequality.) We then have three independent parameters for describing inflation models: $P_{\cal R}$, $P_{T}$, and $n$. Equivalently, we can use normalization $P_{\cal R}$ and the two slow-roll parameters $\epsilon$ and $\eta$. This will prove to be the more convenient choice for categorizing models. For comparing to observation, it proves convenient to use a different set of three parameters: normalization at the quadrupole $Q_{\rm rms-PS}$ (equivalent to $P_{\cal R}$), spectral index $n$, and tensor-to-scalar ratio $r$.
The ratio $r$ of tensor to scalar modes {\it measured at the quadrupole} is\cite{turner93}
\begin{equation}
r \equiv {C_2^{\rm Tensor} \over C_2^{\rm Scalar}} =  13.7 \epsilon,
\end{equation}
so that tensor modes are negligible for $\epsilon \ll 1$. It is a simple matter to transform between slow-roll parameters $\epsilon$, $\eta$ and the parameters $n$ and $r$.

\section{Inflationary Zoology}

Even with the restriction to single-field, slow-roll inflation, the number of models in the literature is large. It is convenient to define a general classification scheme, or ``zoology'' for models of inflation. we divide models into three general types: {\it large-field}, {\it small-field}, and {\it hybrid},  with a fourth classification, {\it linear} models, serving as a boundary between large- and small-field. A generic single-field potential  can be characterized by two independent mass scales: a ``height'' $\Lambda^4$, corresponding to the characteristic vacuum energy density during inflation, and a ``width'' $\mu$, corresponding to the characteristic change in the field value $\Delta \phi$ during inflation.  The height $\Lambda$ is fixed by normalization, so the only remaining free parameter is the width $\mu$.  Different classes of model are distinguished by the value of the second derivative of the potential, or, equivalently, by the relationship between the values of the slow-roll parameters $\epsilon$ and $\eta$. These different classes of models have readily distinguishable consequences for the CMB. Figure 1 shows the $r\,-\,n$ plane divided up into regions representing the large-field, small-field and hybrid cases, describe in detail below.

\begin{figure}
\psfig{figure=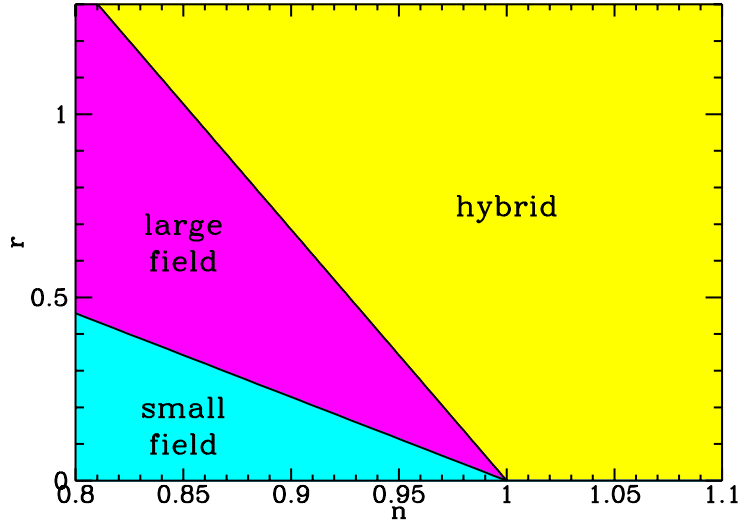,height=3.0in}
\caption{The parameter space divided into regions for small-field, large-field and hybrid models. The linear case is the dividing line between large- and small-field.}
\end{figure}

\subsection{Large-field models: $0 < \eta \leq \epsilon$}

Large-field models are potentials typical of ``chaotic'' inflation scenarios, in which the scalar field is displaced from the minimum of the potential by an amount usually of order the Planck mass. Such models are characterized by  $V''\left(\phi\right) > 0$, and $0 < \eta \leq \epsilon$. The generic large-field potentials we consider are polynomial potentials $V\left(\phi\right) = \Lambda^4 \left({\phi / \mu}\right)^p$,
and exponential potentials, $V\left(\phi\right) = \Lambda^4 \exp\left({\phi / \mu}\right)$. For the case of power-law inflation, $V\left(\phi\right) \propto \exp\left({\phi / \mu}\right)$, the slow-roll parameters are
\begin{equation}
\epsilon = \eta \propto \left(M_{Pl} / \mu\right)^2,
\end{equation}
and $r = 7 (1 - n)$. This result is often incorrectly generalized to all slow-roll models, but is in fact characteristic {\it only} of power-law inflation. Note that we have a one-parameter family of models, parameterized by the width $\mu$.  For chaotic inflation with a polynomial potential $V\left(\phi\right) \propto \phi^p$, the slow-roll parameters are
\begin{equation}
\epsilon = {p \over p + 200} = \left({p \over p - 2}\right) \eta.
\end{equation}
Again we have $r \propto 1 - n$, so that tensor modes are large for significantly tilted spectra. Unlike the case of the exponential potential, the scale $\mu$ drops out of the expressions for the observables, and the models are parameterized by the {\it discrete} exponent $p$.

\subsection{Small-field models: $\eta < 0 < \epsilon$}

Small-field models are the type of potentials that arise naturally from spontaneous symmetry breaking. The field starts from near an unstable equilibrium and rolls down the potential to a stable minimum. Small field models are characterized by $V''\left(\phi\right) < 0$ and $\eta < 0 < \epsilon$. The generic small-field potentials we consider are of the form $V\left(\phi\right) = \Lambda^4 \left[1 - \left({\phi / \mu}\right)^p\right]$. In all cases, the slow-roll parameter $\epsilon$ is very small, so that tensor modes are negligible. The cases $p = 2$ and $p > 2$ have very different behavior. For $p = 2$, the slow-roll parameter $\eta$ is
\begin{equation}
\eta \propto - (M_{Pl} / \mu)^2,
\end{equation}
so that the width of the potential $\Delta \phi \sim \mu$ must be of order $M_{Pl}$ for the spectral index $n \simeq 1 + 2 \eta$ to satisfy observational constraints. As in the case of the exponential potential, the $p = 2$ small-field case is a one-parameter family of models. For $p > 2$, 
\begin{equation}
\eta \propto - \left({p - 1 \over p - 2}\right)
\end{equation}
is {\it independent} of $M_{Pl} / \mu$, so that $\Delta \phi \sim \mu \ll M_{Pl}$ is consistent with a nearly scale-invariant scalar fluctuation spectrum.

\subsection{Hybrid models: $0 < \epsilon < \eta$}

The hybrid scenario frequently appears in models which incorporate inflation into supersymmetry. In a hybrid inflation scenario, the scalar field responsible for inflation evolves toward a minimum with nonzero vacuum energy. The end of inflation arises as a result of instability in a second field. Hybrid models are characterized by $V''\left(\phi\right) > 0$ and $0 < \epsilon < \eta$. We consider generic potentials for hybrid inflation of the form $V\left(\phi\right) = \Lambda^4 \left[1 + \left({\phi / \mu}\right)^p\right].$ The field value at the end of inflation (and hence $\phi_{N = 50}$) is determined by some other physics, and we treat $\phi_{N = 50}$ in this case as a freely adjustable parameter. The slow-roll parameters are then related as
\begin{equation}
{\eta \over \epsilon} = {2 \left(p - 1\right) \over p} \left({\mu \over \phi_{N = 50}}\right)^p
\end{equation}
The distinguishing feature of many hybrid models is a {\it blue} scalar spectral index, $n > 1$. This corresponds to the case $\eta >> \epsilon$, or $\phi_{N = 50} \ll \mu$. Because of the extra freedom to choose the field value at the end of inflation, hybrid models form a {\it two}-parameter family. There is, however, no overlap in the $r\,-\,n$ plane between hybrid inflation and other models. 

\subsection{Linear models: $\eta = - \epsilon$}

Linear models, $V\left(\phi\right) \propto \phi$, live on the boundary between large-field and small-field models, with $V''\left(\phi\right) = 0$ and $\eta = - \epsilon$.

\section{Parameter Estimation from the CMB}

Observations of the CMB don't directly measure $r$ and $n$. What is actually measured is anisotropy in the temperature of the CMB as a function of angular scale. It is convenient to expand the temperature anisotropy on the sky in spherical harmonics:
\begin{equation}
{\delta T(\theta,\phi) \over T_0} = \sum_{l=0}^\infty
\sum_{m=-l}^l a_{lm} Y_{lm}(\theta,\phi) 
\end{equation}
where $T_0=2.726^\circ K$ is the average temperature of the CMB today.
Inflation predicts that each $a_{lm}$ will be Gaussian distributed
with mean zero and variance $C_l \equiv\langle \vert a_{lm} \vert^2 \rangle$.  
For Gaussian fluctuations, the set of $C_l$'s completely characterizes the fluctuations. The spectrum of the $C_l$'s is in turn dependent on cosmological parameters -- $\Omega_0$, $H_0$, $\Omega_{\rm B}$, and so forth. The dependence on parameters is complicated, and  a spectrum of $C_l$'s for a given set of cosmological parameters is calculated by numerically evaluating a Boltzmann equation\cite{seljak96}. Given a set of observational uncertainties $\Delta C_l$ in the CMB spectrum, the goal is to determine observational uncertainties in  $n$ and $r$. Given a set of parameters $\left\lbrace\lambda_i\right\rbrace$, assume a true set of values $\lambda^{\rm true}$. The {\it Fisher information matrix}, given by 
\begin{equation}
\alpha_{ij} = \sum_{l} \left({\partial C_l \over \partial \lambda_i}\right)_{\lambda = \lambda^{\rm true}} {1 \over \left(\Delta C_l\right)^2}  \left({\partial C_l \over \partial \lambda_j}\right)_{\lambda = \lambda^{true}},
\end{equation}
characterizes the accuracy with which the parameters $\lambda_i$ can be measured, with the typical error in the $i^{th}$ parameter being of order $\sqrt{\left(\alpha^{-1}\right)_{ii}}$. The parameters we marginalize over are the spectral index $n$, the tensor to scalar ratio r, the  normalization $Q_{\rm rms-PS}$, baryon fraction $\Omega_{\rm B}$, and the Hubble constant $H_0$. Fixed parameters are the matter fraction $\Omega_0 = 1$ and the cosmological constant $\Omega_\Lambda = 0$. We consider central values for the parameters
\begin{equation}
\left\lbrace\lambda^{\rm true}\right\rbrace = \left\lbrace n,r,Q_{\rm rms-PS},\Omega_{\rm B},H_0\right\rbrace = \left\lbrace 0.9, 0.7, 18{\rm \mu K},0.08,50\right\rbrace.
\end{equation}
The expected uncertainty in the $C_l$'s for a given experiment can be estimated as\cite{knox95}
\begin{equation}
\Delta C_l = \sqrt{2 \over 2 l + 1} \left[C_l + \sigma^2_{\rm pixel} \Omega_{\rm pixel} \exp\left(l^2 \sigma^2_{\rm beam}\right)\right],
\end{equation}
where $\sigma^2_{\rm pixel}$ is the rms pixel nose, $\Omega_{\rm pixel}$ is the area of a pixel in steradians, and $\sigma^2_{\rm beam}$ is the width of the beam, which is assumed to be gaussian. For MAP, we take $\sigma_{\rm beam} = 0.425 \times 0.3^\circ$ and $\sigma_{\rm pixel}^2 \Omega_{\rm pixel} = \left(35{\rm \mu K}\right)^2\left(0.3^\circ\right)^2$. For Planck, we take  $\sigma_{\rm beam} = 0.425 \times 0.167^\circ$ and $\sigma_{\rm pixel}^2 \Omega_{\rm pixel} = \left(3{\rm \mu K}\right)^2\left(0.167^\circ\right)^2$. Figure 2 shows the generic inflation models plotted in the $r\,-\,n$ plane with typical $2\sigma$ error ellipses from MAP and Planck.

\begin{figure}
\psfig{figure=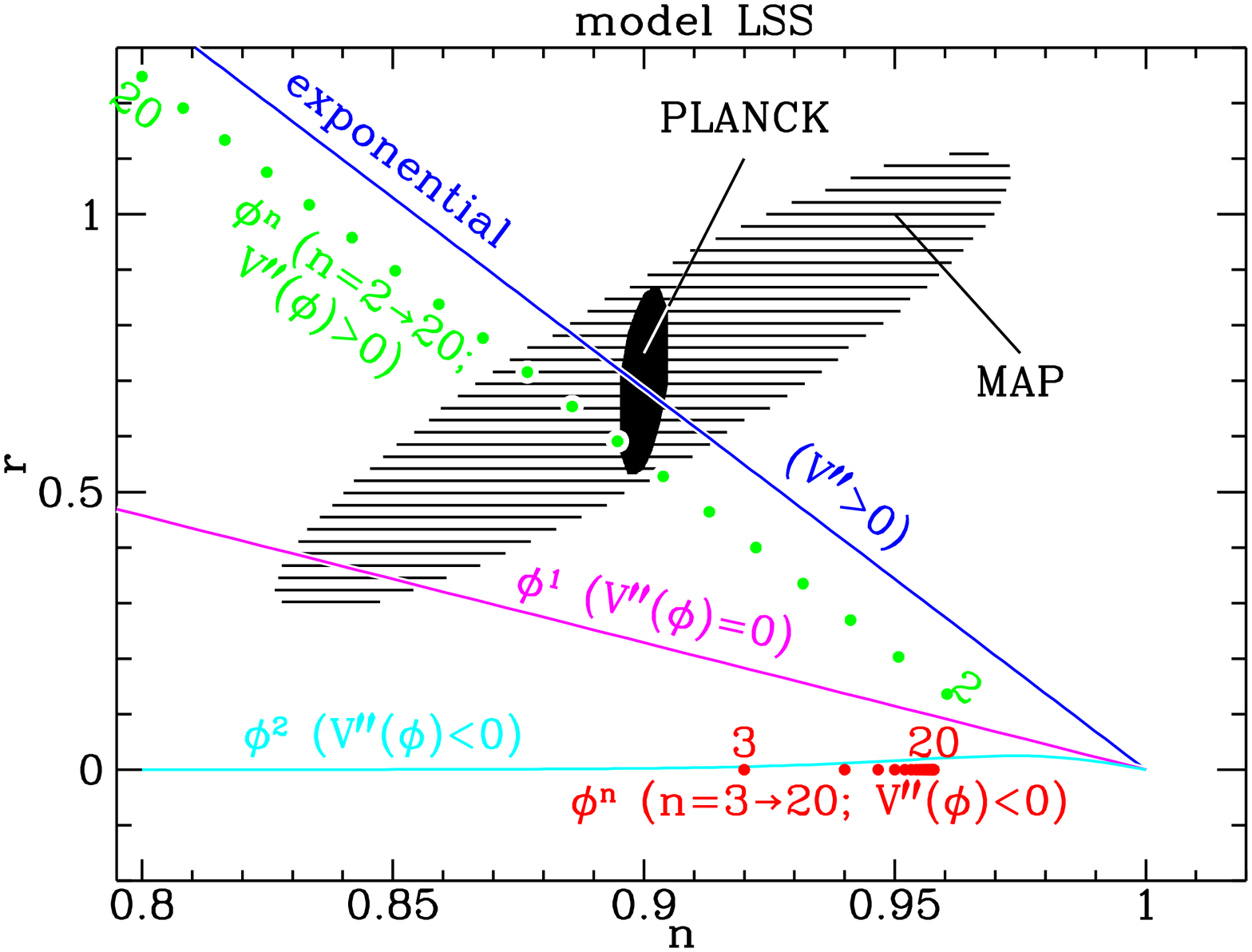,height=3.0in}
\caption{Inflation models plotted in the $r\,-\,n$ plane, with $2\sigma$ error ellipses from MAP and Planck.}
\end{figure}

\section{Conclusions}

It is evident from Figure 2 that MAP will allow at least rough distinction between large-field, small-field, and hybrid models. Planck, on the other hand, will allow for the beginnings of real precision work in constraining inflation models. Planck will be capable, for instance, of distinguishing the exponent $p$ in a chaotic inflation model $V\left(\phi\right) \propto \phi^p$. David Lyth has presented arguments that an appreciably large $r$ is theoretically disfavored.\cite{lyth97} We finish by noting that these arguments will be put to observational test within the next few years!

\section*{Acknowledgments}

We thank Uros Seljak and Matias Zaldarriaga for use of their CMBFAST code. This work was supported in part by DOE and NASA grant NAG5-7092 at Fermilab.


\section*{References}
\begin{thebibliography}{99}

\bibitem{dodelson97} For details and additional references, see S. Dodelson, W. H. Kinney, and E. W. Kolb, {\it Phys. Rev. D} {\bf 56}, 3207 (1997).

\bibitem{MAP} http://map.gsfc.nasa.gov/

\bibitem{planck} http://astro.estec.esa.nl/SA-general/Projects/Planck/

\bibitem{turner93} M. S. Turner, M. White and J. E. Lidsey, {\it Phys. Rev. D} {\bf 48}, 4613 (1993).

\bibitem{lyth97} D. H. Lyth, {\it Phys. Rev. Lett.} {\bf 78}, 1861 (1997).

\bibitem{seljak96} U. Seljak and M. Zaldarriaga, astro-ph/9603033.

\bibitem{knox95} L. Knox, {\it Phys. Rev. D} {\bf 52}, 4307 (1995).


\end{thebibliography}
\end{document}